# Achieving and Managing Availability SLAs with ITIL Driven Processes, DevOps, and Workflow Tools

**James J. Cusick, PMP**
CSO, Director & IT Services Tower Lead for GRC IT
Global Business Services, Wolters Kluwer, New York, NY
j.cusick@computer.org

*Abstract*— System and application availability continues to be a fundamental characteristic of IT services. In recent years the IT Operations team at Wolters Kluwer's CT Corporation has placed special focus on this area. Using a combination of goals, metrics, processes, organizational models, communication methods, corrective maintenance, root cause analysis, preventative engineering, automated alerting, and workflow automation significant progress has been made in meeting availability SLAs (Service Level Agreements). This paper presents the background of this work, approach, details of its implementation, and results. A special focus is provided on the use of a classical ITIL view as operationalized in an Agile and DevOps environment.

*Index Terms*—System Availability, Software Reliability, ITIL, Workflow Automation, Process Engineering, Production Support, Customer Support, Product Support, Change Management, Release Management, Incident Management, Problem Management, Organizational Design, Scrum, Agile, DevOps, Service Level Agreements, Software Measurement, Microsoft SharePoint.

## I. Introduction

For any company running a computing infrastructure and software applications an assumption and requirement is that those systems and applications are available at the time that users and customers need them. In years past systems often ran during business hours but not "overnight". However, in the current environment of global business and Internet commerce there really is no overnight anymore. What might be night time in one city or country is morning or noon time in another. Thus the requirements for availability have been increasing along the dimension of market or regional access. In addition, the demands for reliable computing availability or uptime has also been increasing. With major web sites available around the clock and around the world without any downtime the standards for uptime have forced themselves upward even for internal applications and not just customer facing systems.

For CT Corporation these increasing demands around system and application availability are no different. Over the last 10 years CT has made increasing investments in the engineering, process support, and team strength to meet and manage defined SLAs (Service Level Agreements) across each application in its portfolio. Starting in 2008 the CT Operations team began putting in place the practices, tools, processes, engineering, communications, and organizational approaches required to boost availability and meet or surpass our negotiated SLAs [1]. With rare exceptions, we have been able to achieve this. In this paper, we will detail the origins of this work, the incremental progress made to achieve these results, the specifics of some of the methods, and the outline of the quantified results.

But first a brief introduction of Wolters Kluwer. Wolters Kluwer (WK) is a €4.3B Netherlands-based international publisher and digital information services provider with operations around the world. Wolters Kluwer is organized into Divisions and Business Units. The experience documented here focuses on work done for New York-based CT Corporation (CT). The CT IT Operations team is a part of Wolters Kluwer's Global Business Services organization dedicated to supporting the Governance, Risk, and Compliance (GRC) Division.

The systems operated by CT include public-facing Web-based applications and internally used ERP (Enterprise Resource Planning) systems. Major vendors manage network services and hosting for our computing environments. CT IT is responsible for the development and operations of these systems from an end customer standpoint. It is the CT IT Operations team which has been primarily responsible for the availability processes and tools documented here although multiple parties contributed to the achievement of these results.

The systems and applications SLA management work reflected here benefits our customers directly by providing acceptable uptimes and issue response times for our systems. They also help our development organization by isolating troubleshooting and problem resolution to allow them to focus a greater portion of their time on new product development. This work also reduces costs, improves flexibility, and speeds system evolution by allowing for rapid adaptation in the face of technical or process issues.

## II. AVAILABILITY BACKGROUND

Within computing there are several core terms associated with the consistent delivery of IT system and application services. Availability is defined as the percent of uptime during planned uptime windows or schedules. Driving availability is the reliability of the system and its supporting operations processes. This can be further broken down into the reliability of the composite applications or components.

To be specific about some of these terms let us present some essential definitions. Availability is defined formally, as mentioned, as the percent of uptime within planned operational windows and is calculated as such [2]:

*Availability = Uptime / (Uptime + Downtime)*

Thus, it is important to know both the planned schedule of system operations as well as the realized downtime within that window to compute availability.

A key factor leading to availability is reliability. Reliability is defined as the probability of failure free operation in a given time period [3]:

$R(t) = e^{\wedge}(-\lambda t)$

where
  R = Reliability
  *t* = time period and
  *λ* = failure intensity rate

Finally, when required, to convert from reliability to availability (moving from a probability to a percentage) simply multiplying by 100 is the approach:

$R(t), \% = (e \wedge -\lambda t) * 100\%$

In developing availability SLAs the traditional approach is to construct an uptime target as a percent of the total planned operations window. It is generally assumed that most systems cannot achieve 100% uptime indefinitely as some component will fail eventually unless every component has a high-availability and redundant failover capability and/or a fault-tolerant design which might include dynamic functional re-routing, tertiary backup systems, or other methods. Such designs are often found on spacecraft, aero-space systems, aircraft, and other life critical systems [4]. As a result, for most systems a sliding scale of availability as a percent of uptime (or the number of "9's") is often considered. This scale is presented in a ladder organized by orders of magnitude of availability as in *Table 1* [2]. This table also conveniently shows exactly how much downtime is allowable per SLA level.

Within CT the specific SLAs selected per application is drawn from this range of target values. The decision to select one of these availability goals is based on a number of factors including product direction, target user population, economics around design, maintenance, and support, and other factors. In this paper we will be discussing the process, tools, and methods supporting our SLA achievement and providing less detail around the SLAs targets themselves. We believe that the methods we describe are well proven and can be utilized in many other environments.

| Availability % | Downtime per year | Downtime per week |
|---|---|---|
| 99% ("two nines") | 3.65 days | 1.68 hours |
| 99.9% ("three nines") | 8.76 hours | 10.1 minutes |
| 99.99% ("four nines") | 52.56 minutes | 1.01 minutes |
| 99.999% ("five nines") | 5.26 minutes | 6.05 seconds |

*Table 1 – Typical Availability SLA Categories*

## III. EVOLUTION OF THE PROGRAM

### A. Foundational Components

Beginning in 2008 CT started organizing its support function under one department where it had earlier been separated across groups. What eventually became the CT Operations Center was born. Over the years there have been some groups added and others realigned but the mission to provide system and application availability meeting negotiated SLAs has been a constant. Within the operations function there are core groups consisting of an Operations PMO (Project Management Office), Product Support, Business Analysis Support, Engineering Support, Application Support, and core infrastructure support. These teams are primarily "in-house" but a small number of key vendors also participate especially in the areas of hosting and offshore technical support. Organizational coordination is provided through a core management staff along with partner arrangements for vendors and extended teams.

This organization continues to adapt itself to the challenges it faces. In recent years, the center created a Level 2 Business Analysis function and also created a special purpose "Technical Services Desk" or TSD to handle complex system issues at the Level 1 desk. Finally, in the most current organizational change, the Operations Center was aligned with the Wolters Kluwer Global Business Services organization and continues its support for the GRC Division and CT Corporation from there. This change has not modified the processes described below to deliver on our SLAs. However, it is expected that with time the broader organization will allow for the achievement of improved processes in support of these availability goals.

Some of the key ITIL (Information Technology Infrastructure Library) oriented areas [5] which the organization has focused on to achieve and manage availability SLAs include:

1. Architecture and Design
2. Monitoring and Alerting
3. Release Management
4. Change Management
5. Incident Management
6. Problem Management
7. Production Support with Scrum
8. Capacity Management
9. Security Operations
10. Financial Management

We will now describe the work done in each of these areas and how they relate to SLA achievement. No single area can deliver success in reaching a specific SLA target. However, the balanced and synchronized management of each of these disciplines can result in a successfully attaining the level of performance desired.

*B. Architecture and Design*

Naturally, the reliability and therefore the availability of any system is directly related to its architecture, design, and implementation quality. For CT, nearly all systems are architected with a level of high-availability and fault tolerance depending upon requirements. This is typically achieved by using redundant servers at each architectural tier and often some level of data replication or simultaneous data commit. In some cases, platforms with the highest SLAs have significant real-time failover capability designed-in from the start using such technologies as *RAC* (Real Application Clustering) from Oracle or equivalent technologies from competing vendors.

Beyond server and database redundancy at the design level CT also develops applications which provide intelligent sustainability in the face of errors. For example, if a primary messaging path is not available, applications are sufficiently intelligent to queue requests and retry when the connection is available thereby preventing key customer transaction information from being lost. Such designs are fundamental to attaining highly available systems. Beyond this architecture, we enter the realm of operations which then helps assure that these implementations can live up to the designer's intent in production. It is often through operational experience that improvements to design approaches are fed back into the architecture process for future solutions.

*C. Monitoring and Alerting*

With an architecture in operations, each system, application, and component requires the appropriate and effective level of monitoring and alerting as a first line of defense to achieving a given SLA. CT utilizes three broad monitoring and alerting mechanisms.

1. The first monitoring toolset is applied at the sever and component level within the datacenter. This set of monitoring is applied to each server and critical subcomponent. This can include the CPU, memory, disk, storage, and network characteristics and are typically created using a tool like *BMC Patrol*. This level of monitoring is managed by our hosting vendor and is applied in standard profiles based on the criticality of the device. Thresholds are negotiated and if an alert is triggered an automated support ticket is created. At that point standard incident response protocol is followed as discussed below.
2. In addition to these tools which run internal to the datacenter CT also runs monitoring tools which originate from outside the proprietary network. The primary such tool is provided by *Keynote* a popular tool which allows for a variety of scripted "keep alive" pings from across the public Internet to continuously test application availability. These tools provide both uptime data and also performance (response time) statistics. Should any of these scripts detect a violation against a threshold value an alert is raised. Once again, standard incident response tactics are then employed as described below.
3. Additionally, CT has recently deployed the *NewRelic* toolset to provide APM (Application Performance Monitoring). This is a robust toolset that instruments website performance, availability, and functional usage. This tool is particularly useful to the development team but is also used by the operations team to obtain visibility into application performance and behavior.
4. The final broad category of alerting and monitoring is a set of custom logging and alerting mechanisms built into the applications or added to the system layer. Typical examples include automated database health checks or routine file storage growth rate alarms which provide a custom overlay to the out-of-the-box monitoring tools. In addition, the CT portfolio of applications is built on a shared library of common code routines including a common logging capability. This capability utilizes a collector database which then allows for real-time or after the fact analysis of practically any application page in terms of its behavior, performance, or anomalies encountered.

This combination of monitoring and alerting tools and approaches at different levels within the architecture allow for sufficiently granular automated observation and promotes adequate response time helping drive availability upwards. One key requirement of the monitoring architecture and implementation itself is that it needs maintenance also. For example, when new systems are deployed the monitoring requirements need to be established and implemented. The same is true when there are upgrades or modifications to the infrastructure. Missing this step can leave a critical device without the appropriate monitoring coverage. This step needs to be captured in pre-golive planning. At CT we do this through our Production Readiness Review (PRR) process [6] discussed in more detail below under Release Management.

*D. Release Management*

Building on the areas of architecture and monitoring comes Release Management. Release Management governs the changes that will be deployed to the production environment as they are packaged at the highest level in releases. Change

Management, discussed next, covers the finer grain decisions around changing the production environment but these two areas are closely linked. Within CT's process this function is managed by the Operations PMO team.

The core elements of this process include Product Support/Release Management, Calendar Management, and our change board called the Migration Review Board. Releases are built from selected PBIs (Product Backlog Item). These collections or releases are then scheduled in the non-production environment for testing and simultaneously for projected release to production. This production release scheduling is provisional based on several conditions including QA certification, logistics, release window availability, and release preparation. This includes a Production Readiness Review which reviews a checklist of applicable pre-golive criteria such as storage configuration, firewall rule provisioning, and more.

This process area contributes to the achievement of planned availability SLAs by managing the rate of change and helping to ensure both controlled change introduction as well as well prepared and high quality releases only make it into production.

### E. Change Management

Change Management works hand in hand with Release Management and as a sub-process. Change Management aims to prepare, coordinate, execute, and confirm all changes to a production environment once the meta-releases have been approved. These can be software or hardware changes or even data content or configuration changes. For CT, we actually have a dual layer change management process. The first layer covers changes to our hosted environment as managed by a third party. This requires extensive collaboration and coordination between CT and its hosting partner. This change process is highly regulated and supported by an integrated tool set built on a version of *BMC Remedy*.

The second layer of change management is governed by the CT Operations PMO. This covers all proprietary software and data changes required to support revenue supporting applications residing on the hosted environment. Some of the details of this process as driven by release management include the tagging of specific changes for a particular release, issuing formal requests for these changes to be deployed, recording these requests in our change request repository, reviewing these requests daily, and ensuring that system engineers and database administrators with production permissions only make changes which are formally approved.

Change management thus is a vital process in contributing to availability. It is said that all improvements require a change. Thus, to keep improving, changes need to be done accurately and well.

### F. Incident Management

Incident management is a key process area to help assure high availability if something actually does go wrong in production. It is important to know the time between incidents, how fast an incident is responded to, and how long resolution of an incident takes. Then, each of these activities can be studied to attempt to reduce their intervals. The classic measurements applied are MTTF (Mean Time to Failure), MTTR (Mean Time to Repair), and MTBF (Mean Time Between Failures) [2]. The relationships of these metrics are shown in *Figure 1* [7]. Within the CT availability program each of these metrics are tracked and improvement initiatives have been pursued over time to reduce each interval.

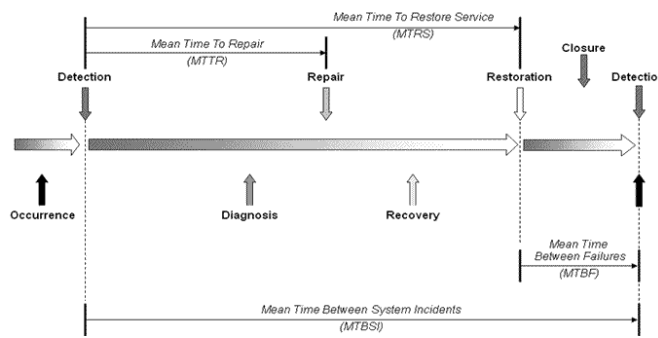

*Figure 1 – The Incident Lifecycle [7]*

The concept of formal Incident Management is well understood in industry standards including ITIL[1] (IT Infrastructure Library). According to ITIL [8]:

> *An 'Incident' is any event which is not part of the standard operation of the service and which causes, or may cause, an interruption or a reduction of the quality of the service.*

The objective of Incident Management is to restore normal operations as quickly as possible with the least possible impact on either the business or the user, at a cost-effective price. Activities of the Incident Management process include:

- Incident detection and recording
- Investigation and diagnosis
- Resolution and recovery
- Classification and initial support
- Incident closure
- Incident ownership, monitoring, tracking and communication

The CT team decided to implement the essential aspects of the ITIL Incident Management model which laid a path for the improvement of service outage management. An "Incident

---

[1] *http://www.itil-officialsite.com/home/home.asp*

Response Team" was created which is referred to as the IRT. The IRT covers nearly all aspects of Incident Management in a light weight fashion. This team was later put under the management of the support organization with the creation of a new overlay team called the TSD (Technical Services Desk).

There were several key objectives to the creation of our incident management process including providing standardized coverage around the clock, incident logging, management reporting, and efficient and effective incident follow up and closure. We have achieved all of these goals and more by developing this ITIL inspired process [1]. We also implemented a custom ticketing tool using Microsoft's *SharePoint* environment to support the process. While there are many commercial tools of this type we have found this approach to be highly flexible and extensible especially since the entire implementation was under our control. Currently Wolters Kluwer is in the process of implementing *ServiceNow* globally, thus we will in all likelihood be transitioning to that application in the near future.

In terms of incident follow through, for each incident an analysis is done to determine if it warrants a problem management analysis based on its severity and impact. Similarly, when incidents generate outages and the incident is of high severity an availability record is created via workflow. This record is then reviewed with the facts of the outage and the duration of the outage and systems impacted are confirmed and recorded. Once this is done a background job refreshes a graphical dashboard from the availability record database implemented in Microsoft's *SharePoint*. The dashboard generation logic is implemented in C# and .Net. The dashboard displays the applications in CT's portfolio and their near real-time availability achievement. Several views are available for the user within this custom availability SLA dashboard including a per product view and a toggle between availability as a percent of uptime and cumulative minutes of downtime which is sometimes more understandable for business stakeholders (see *Figure 2*). This availability data also exposes an API which some downstream systems use. In one case this data is used to help manage the delivery SLAs of vendors. Essentially, if outage thresholds from this process are reported in excess of negotiated SLAs then contractual penalties can be applied.

Recently an operational performance dashboard was developed in order to brief the Executive stakeholders on the trends and impacts of these events and incidents as well as the operational performance as relates to predefined SLAs. This dashboard draws on data related to incident volumes, types, durations, and business impact. In addition, the Executive dashboard also includes data around user support cases, defects, and the rate of closure of these cases. Importantly, this data is analyzed by the application owners to develop themes around key operational challenges and what approaches will be taken to mitigate them. This level of management involvement in regularly reviewing and discussing these operational results and performance against expectations has gone a long way to focus effort on the highest impact areas and prioritize solutions. This dashboard employs a rich set of data visualization techniques and at the same time compresses information into a compact and readable format for quick comprehension. This dashboard continues to evolve in its content and in its use.

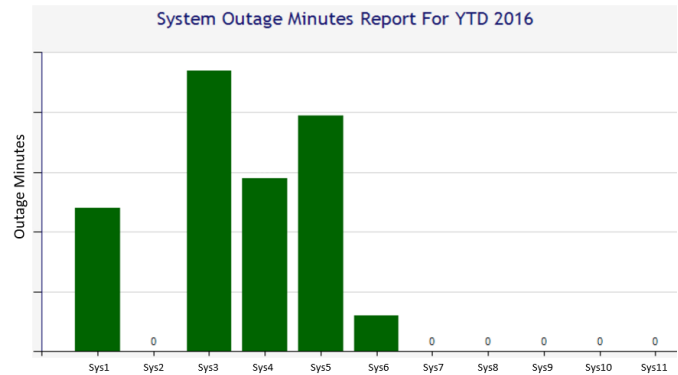

*Figure 2 – Sample Outage Minutes Dashboard Report*

### G. Problem Management

The primary goal of problem management is to eradicate the root causes of any incidents. This ensures they do not repeat and thereby begins building up availability further. There is a complex relationship between incidents and problems. Typically, a problem will cause an incident and that incident will be generated each time the underlying problem is encountered. In some cases, a problem can cause different types of incidents and in rarer conditions more than one problem can cause what appears to be an incident with the same characteristics. Problem management is all about cutting through this complexity and "rooting out" the true causes.

CT's process includes the automatic creation of a problem ticket from any significant incident event. The automatic creation of a problem ticket includes the assignment of the ticket to a pre-identified resolver and their management chain. A 10-day SLA is in place for the assignee to present their analysis in a documented fashion within the *SharePoint* based system. Finally, an independent reviewer of the RCA (Root Cause Analysis) report is included to prevent any rush to conclude the process prior to obtaining a full and complete exploration of the problem along with proactive measures to prevent reoccurrence of the problem.

The process is supported by a methodology document with defined inputs/outputs, procedural steps, and automated workflow steps. Furthermore, analytical tools provide by the Taguchi Method (aka, "fishbone diagrams") [9] and the 5 Why's method [10] are supplied in the process approach artifacts. Additionally, sample RCAs from prior problems are included in the repository to assist assignees and participants in the root cause process to attain a rich analytical result. These templates include timelines, fishbone diagrams, and clear conclusions around the causes and suggestions for preventative and proactive steps.

While the process itself is robust and is followed, especially in the case of a significant incidents, we have found that with the pace of change and project deadlines sometimes the RCA analysis work receives less priority. We continue to look at mechanisms to make the process easier to follow and to reward participants working within the process. One of the biggest challenges with conducting RCAs is that each one is somewhat unique and requires detailed and time consuming review, analysis, and creative solution development. It remains difficult to put a time frame around this type of work in many cases.

*H. Production Support and Scrum*

While infrastructure issues contribute to many incidents thus negatively impacting availability within our environment, the predominant cause of incidents can be traced back to application faults. Naturally there are many approaches to reducing the number of software defects released to production to thereby improve software reliability. However, for the purposes of this discussion it is worth pointing out that a transformation effort our development teams have undertaken in recent years has been a key factor in improving software quality and availability especially in green-field application development efforts. In specific, an adoption of Agile methods and Scrum have significantly impacted application quality. While the legacy systems have not benefited as much from these new methods the new applications which we have brought to market have. Over time we expect these improvements to extend to all applications in our portfolio.

*I. Security*

Similar to the CT incident management process is our Computer Security Incident Management Team process (CSIRT) which is defined as part of our Information Security program [11]. Our InfoSec program provides a governance model, a security controls framework, policies, procedures, a Secure SDLC, and supporting tools and processes. Within the ITIL framework security is a key component. Our defined InfoSec program fulfills the requirements of this area. A major effort today is scaling out this program to aligned Business Units and incorporating the program into our DevOps (or DevSecOps) transition approach as discussed below. Our ultimate goal with the InfoSec program is to embed the needed steps into each area of the organization as a routine and integrated set of capabilities and not as something seen as driven externally by our security team. This is the promise of the DevSecOps model where all players in the development and operations functions work closely together including security.

IV. DEVOPS TRANSFORMATION

Beginning in 2015 our operations team began taking an in depth look at what the concepts, approaches, and tools of the DevOps community could mean for our work. We conducted research into the subject and invited a consultant to facilitate a workshop on DevOps adoption [12]. This workshop brought together product management, development, QA, support, operations, and management. Over the several days of consultation and the time that have followed we developed a set of priority areas to focus on by simply asking the question – "what could be improved with our current approaches". Some of the answers were around collaboration models, deployment automation, test automation, continuous integration, better use of customer support data in product development, and more.

Today our transformation to a broader use of DevOps remains a work in progress but it is a significant goal and focus area. Our development teams invested in Scrum based Agile development over the last few years and have reaped noticeable rewards. From an operations perspective, our most significant progress thus far has been in automating both builds and deployments. This covers both our code base and our database releases. We have actually seen deployment times drop from hours to minutes in certain areas. This provides us with significant new options and agility in development planning and operations.

We believe DevOps or DevSecOps will also provide significant benefits in our general operational functions. Blending our security operations into our operational work is critical in meeting many of our customer's expectations and such approaches have been shown as beneficial across a set of capabilities [13]. One thing we believe is critical, however, is to judiciously convert our IT service catalogs to processes within the DevOps model that continue to deliver on the SLAs that we have been delivering on for years. We cannot simply start from scratch but need to leverage the processes and tools we have in place today and shape them anew within a DevOps model.

Since we have a mature operations practice we believe we can do this gracefully. One of the inspirational texts we have been applying is "The Phoenix Project" [14]. This highly popular novel about applying DevOps has provided our staff with a light-hearted view into the types of trials and tribulations they often face in developing and supporting IT systems. Importantly, it has actually already motived our team to define a set of improvement goals in broad areas of process, technology, and communication. We look forward to seeing how these empowered teammates will take us to the next level in applying DevOps and continuing to reach our SLAs.

## V. OPERATIONAL EXPERIENCE

### A. Initial Experiences

The foundations of our availability program were put in place in 2008. At that time, the focus was on establishing release management, change management, and incident management. We defined these practices, educated stakeholders and other required participants, and tuned the processes and tools. We saw immediate improvements in availability due to strong controls around releases and better MTTR due to tighter incident management. We also realized a 40% lengthening in MTTF between 2009 and 2010. Perhaps the most significant benefit at this stage was gaining quantified controls around the system behaviors and support operations. Some of the challenges we faced in the initial years centered on establishing necessary management and business support, buy-in from all technical participants, and especially building out adequate staffing coverage schedules.

Another key challenge we sometimes still face is a lack of understanding of the processes especially from people new to the organization or those taking on support functions for the first time. Quite often we need to repeatedly explain the processes so that people understand their purpose, flow, and implementation. This is one area that requires continued patience as it is understandable that people will question things which they are not familiar with or where they believe they can suggest improvements even if they do not have a full understanding of a debugged and operational process. Of course, we always welcome suggestions for improvement and have over the years even revamped parts of the process as warranted.

### B. Process Maturation

As the processes developed focus also expanded into problem management, availability reporting, capacity management, and security. Perhaps the most successful of these areas have been availability reporting and security. As mentioned earlier we implemented a real-time availability dashboard which is available online to anyone in the company. This provides excellent feedback to application teams on their production results against defined SLAs. As for security, CT has focused on its Information Security program intensively over the last three years to supplement the corporate level security program which has been in place indefinitely. The work in this area contributes to availability attainment by reducing security related impacts to production systems [11].

The areas where we have had challenges have been in problem management and capacity planning. With problem management, we have a robust process and supporting toolset. However, there has been ongoing issues in prioritizing RCAs over other work. Typically, application teams only conduct detailed RCAs on the highest impact problems. However, this has been improving over the last year or two. As for capacity planning we have attempted to build a capacity planning and forecasting model and process but this has not come to full maturity yet. Furthermore, with the advent of Cloud Computing, the use of formal capacity plans is beginning to wane. Since Cloud Computing provides for dynamic scaling forecasting becomes much less of an issue. Perhaps this only becomes an economic forecasting issue in the future and not as much of a computing resource issue as in years past.

### C. DevOps Adoption and Process Conversion

As a result of some of the above challenges we are hoping to leverage our work in DevOps transformation to help integrate all participants in the processes driving SLA achievement around availability more fully. We believe that a DevOps mentality and approach as well as commonly used tools from the DevOps community will help in ensuring full engagement of all participants covering development, QA, operations, and support in defining best in class approaches for maintaining availability. We are hopeful that instead of development viewing the support processes as ones imposed on them that they will see them as ones they helped to define and agree with at a fundamental level.

## VI. OBSERVATIONS, LIMITATIONS, AND FUTURES

The overall approach CT takes to ensuring SLA attainment is a traditionalist one which values stability for our customers over rapid an unmeasured change. To be clear, we release software and make infrastructure changes regularly and have the ability to respond to emergency changes within this process framework. However, the bias is toward measured change and stability. This is required due to the mission critical nature of our systems and the fact that our customers and our business count on their reliability and resultant availability to meet their operational needs.

However, we have adopted a bi-fricated model in recent years where new systems or beta releases get a different or more nuanced treatment around risk. If a new application is in beta or limited introduction for example, we might treat its release criteria differently than a change to one of our mature mission critical systems. This is what Gartner calls Bi-modal IT [15]. We do subscribe to this model and are able to flex as required. Furthermore, as we adopt DevOps methods and automation more and more the speed of changes via Continuous Integration and Deployment will only ramp up and our process will need to evolve to meet these requirements.

## VII. CONCLUSIONS

Our journey toward higher SLA attainment has not always followed a straight line. At times, we have made progress and then we have seen some setbacks. What we have found, however, is that the essential and foundational process areas of release, change, and incident management brought quick order to our environment. Once these processes were in place and communicated we could then iterate on them, strengthen them, and expand into additional supporting process areas. This was also quickly supported using COTS (Commercial-of-the-Shelf) tools and also by our development of custom tools to bring these processes to life and to enforce the core requirements of these processes. Importantly, improving the skills and shared understanding of our staff around the goals of these processes continues to be essential to our progress.

Finally, the emergence of the DevOps model of thinking about development and operations working together to achieve higher quality, faster turnaround time, automation, and joint understanding from each of the product, development, QA and operations team's perspectives and specific requirements has led to significant further improvements to our SLA management and achievement especially for green-field applications. We remain optimistic that our work in this area will help us find new areas of innovation in achieving stronger levels of operational success for our customers and we believe that the approaches described here can have wide applicability in the industry.

## VIII. ACKNOWLEDGEMENTS

The development of CT's availability management approach and environment developed over many years and a wide variety of people contributed to its maturation. The author is particularly thankful to the contributions of Gary Ma, David Miller, Nick Laurita, Susan Claudio, and the entire support team. In addition, dozens of developers, managers, engineers, technicians, analysts, project managers, contractors, and product managers followed the process framework day in and day out and provided feedback towards its improvement. In some notable cases, it was their process designs or observations which pushed the effectiveness of the approach forward beyond what the author would have imagined. Finally, special thanks to Murugan Gnanavel wrote custom app which provides the real-time Availability Dashboard.